\documentclass[12pt]{article}
\usepackage[a4paper,margin=1in]{geometry}
\usepackage{lineno}
\usepackage{booktabs}
\usepackage{multirow}
\usepackage{float}
\usepackage{authblk}
\usepackage{amsmath}
\usepackage[pdftex]{graphicx}
\usepackage{hyperref}

\title{Experimental insights into data augmentation techniques for deep learning-based
multimode fiber imaging: limitations and success}

\author[1]{Jawaria Maqbool}
\author[1]{M.~Imran Cheema\thanks{Corresponding author: imran.cheema@lums.edu.pk}}
\affil[1]{Department of Electrical Engineering, Syed Babar Ali School of Science and Engineering,\\
Lahore University of Management Sciences, Lahore, Pakistan}
\date{}

\begin{document}

\maketitle

\begin{abstract}
Multimode fiber~(MMF) imaging using deep learning has high potential to produce compact, minimally invasive endoscopic systems. Nevertheless, it relies on large, diverse real-world medical data, whose availability is limited by privacy concerns and practical challenges. Although data augmentation has been extensively studied in various other deep learning tasks, it has not been systematically explored for MMF imaging. This work provides the first in-depth experimental and computational study on the efficacy and limitations of augmentation techniques in this field. We demonstrate that standard image transformations and conditional generative adversarial-based synthetic speckle generation fail to improve, or even deteriorate, reconstruction quality, as they neglect the complex modal interference and dispersion that results in speckle formation. To address this, we introduce a physical data augmentation method in which only organ images are digitally transformed, while their corresponding speckles are experimentally acquired via fiber. This approach preserves the physics of light-fiber interaction and enhances the reconstruction structural similarity index measure~(SSIM) by up to 17\%, forming a viable system for reliable MMF imaging under limited data conditions.
\end{abstract}

%%%%%%%%%%%%%%%%%%%%%%%%%%  body  %%%%%%%%%%%%%%%%%%%%%%%%%%
\section{Introduction}  \label{sec:Intro}
Multimode fiber~(MMF) imaging is a challenging inverse problem in which the object image is transmitted through a multimode optical fiber, resulting in a complex speckle pattern due to modal dispersion and interference. Mainly, three different methods are used to reconstruct original images from these speckle patterns: transmission matrix measurement, phase conjugation, and deep learning. Transmission matrix and phase-conjugation methods require phase measurements via a reference arm, which makes the setups complex~\cite{popoff2010measuring,akbulut2013measurements,papadopoulos2013high,papadopoulos2012focusing}. However, experimental setups can be simplified in the case of deep learning, as only the speckle intensity is measured at the fiber’s output. A neural network is employed to reconstruct the original image from these intensity patterns ~\cite{rahmani2018multimode,borhani2018learning,zhu2021image,kremp2023neural,maqbool2024towards,feng2025high}. As a speckle image is being converted to an original image, this task comes under the image-to-image translation category of deep learning. Deep learning's ability to reconstruct high-fidelity images from speckle patterns has made MMF imaging a promising approach for developing ultra-thin, flexible endoscopes.

%Along with multimode imaging, deep learning has revolutionized science and technology in the last decade. Significant advancements in medical imaging analysis, precision agriculture, remote sensing, and signal processing have been made possible by its development ~\cite{lecun2015deep,esteva2021deep, upadhyay2025deep,lu2025vision}.
The strength of deep learning models in problems such as image classification~\cite{krizhevsky2012imagenet,kumar2024deep}, image-to-image translation~\cite{isola2017image}, segmentation~\cite{brar2025image,liew2017regional}, object detection~\cite{aung2024review}, and super-resolution ~\cite{lee2024holosr,su2025review} stems from their ability to learn complex patterns from a large number of images. As a result, the size and diversity of training data greatly influence model performance in both classification and regression tasks~\cite{sun2017revisiting,russakovsky2015imagenet}. Acquiring large and labeled image datasets is often impractical in domains such as medicine, agriculture, remote sensing, and even in multimode fiber-based endoscopy due to privacy, consent, environmental variability, or logistical constraints~\cite{goceri2023medical,meng2025tlstmf,cieslak2024generating}. To address this problem, data augmentation expands datasets through transformations or synthetic generation, aiming to improve generalization, reduce overfitting, and correct class imbalance~\cite{sharma2025addressing}.

Data augmentation in deep learning is generally achieved through two main approaches: standard image-based augmentation and generative model-based synthesis. Standard augmentation includes geometric operations such as flipping, rotation, translation, and scaling, as well as photometric modifications, including jittering of color or intensity~\cite{de2022geometric}. Researchers have demonstrated that augmentation approaches improve performance in classification, segmentation, and detection tasks~\cite{mumuni2024survey, kumar2024deep}. These improvements become noticeable in scenarios involving limited datasets, class imbalance, or when the training data captures only a limited range of environmental conditions and fails to represent variations in brightness, contrast, orientation, or scale~\cite{goceri2023medical,khalifa2022comprehensive}. However, the performance of data augmentation is context-dependent and varies with the type of imaging procedure. Recent studies have questioned the notion that augmentation is always advantageous, showing that while certain transformations can improve overall accuracy, they may degrade performance for specific classes by adding ambiguity between them~\cite{kirichenko2023understanding}. This occurs when harsh transformations distort or remove critical features~\cite{gong2021keepaugment}. Yet the problem of whether standard data augmentation techniques work or not for multimode fiber imaging remains uninvestigated.

In addition to simple transformations, deep generative models, such as variational autoencoders~(VAEs), generative adversarial networks~(GANs), and diffusion models, are widely used to synthesize realistic and diverse samples. In this work, we focus on a GAN-based approach and standard augmentation techniques. For brevity, VAEs and diffusion models are excluded, as they represent broad research areas that require separate research. Generative models are useful when data availability is limited and acquiring additional data is time-consuming and costly~\cite{yue2022survey,mumuni2024survey}. Various GAN architectures, including CGAN, CycleGAN, and DCGAN, are generally employed to reduce data bias, avoid overfitting, and enhance generalization and accuracy~\cite{ding2025improving,paproki2024synthetic,cieslak2024generating,thakur2025ai}. However, recent works show that GAN-based augmentation does not necessarily improve performance; in some cases, it can even decrease model accuracy and increase data bias~\cite{bissoto2018skin,ali2023leveraging,jindal2024bias}. These findings highlight that while GANs are useful in data-limited domains, their benefits are highly task-specific and depend greatly on the realism and diversity of the generated data. It is important to mention here that although conditional generative adversarial networks~(CGAN) have been previously used for image reconstruction in MMF imaging~\cite{yu2021high,wang2022upconversion,maqbool2024application}, they have not been exploited for data augmentation.

 Like other image-to-image translation tasks, multimode fiber imaging using deep learning can achieve robust, generalizable performance when trained on large and diverse datasets. These datasets should capture the full range of real-world situations, including variations in image orientation, brightness, and contrast. But if we are using these multimode fibers in medical endoscopy, data scarcity remains a significant challenge, limiting their performance. Even with relatively large datasets available, acquiring their speckles remains a major experimental challenge. Each input image must be displayed on a spatial light modulator~(SLM), transmitted through the MMF, and recorded at the output as a speckle pattern. This process is time-consuming and sensitive to environmental changes such as temperature, vibration, or fiber bending. For example, in our experiments using the OrganAMNIST dataset, which contains 58,830 images, capturing the corresponding speckles requires nearly 25 hours~\cite{maqbool2025deep}. Such long acquisition times not only constrain scalability but also pose a risk, as they may lead to data consistency loss due to system drift or instability. This experiment time further increases when the goal is to make multimode fiber imaging robust against fiber perturbations. In such cases, speckle patterns must be recorded for multiple fiber configurations and bending conditions to ensure model generalization and stability for different states~\cite{abdulaziz2023robust}. Together, these constraints motivate several key research questions central to this study:(1) When the available data is limited, can standard augmentation methods, such as flipping or rotating both the label and its corresponding speckle, improve the average structural similarity of reconstructed images? (2) For large datasets, can CGANs be used to generate synthetic speckles, thereby reducing experimental data collection time without compromising reconstruction quality? To the best of our knowledge, a systematic investigation of these research problems has not been addressed in prior literature.

In this work, we experimentally and computationally demonstrate that applying standard image transformations to both label and speckle pairs degrades reconstruction performance. We also demonstrate that CGANs fail to generate realistic, task-relevant speckle patterns; adding their synthetic data to real datasets actually reduces the average SSIM of reconstructed images. In addition, we propose and implement a physical data augmentation strategy exclusively designed for multimode fiber imaging. In this method, only organ images are rotated, cropped, and flipped, and their corresponding speckles are physically recorded using the SLM–fiber system. This augmentation technique improves SSIM notably. We also demonstrate that different transformation types have different impacts on the SSIM value. Our results emphasize that, unlike in conventional deep learning tasks, effective data augmentation in MMF imaging requires physical awareness of the imaging process, with the optical system included in the data augmentation pipeline. By examining when augmentation works and when it fails, this work provides a new approach to improve reconstruction fidelity under practical constraints.

\section{Experimental framework and data collection}  \label{sec:exp}
\begin{figure}[h]
\centering
\includegraphics[width=5.2in]{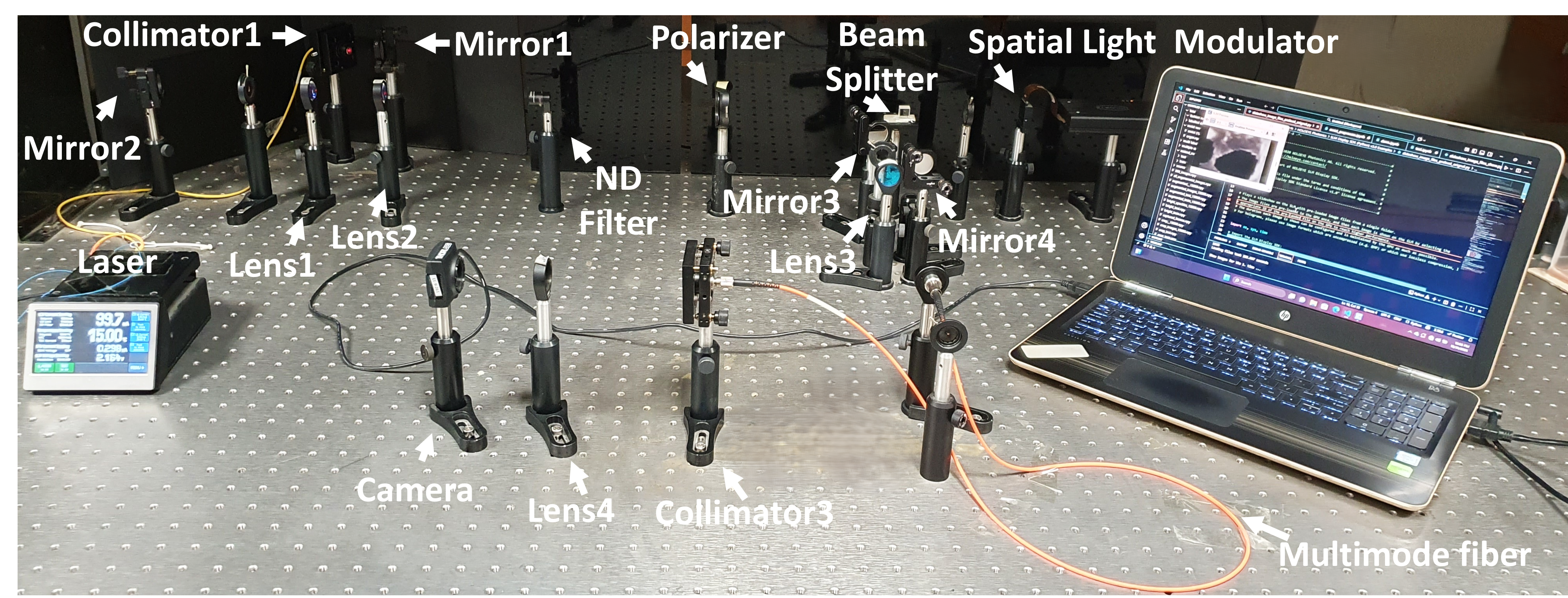}
\caption{Experimental multimode fiber imaging setup showing the complete optical path from the laser source to the camera, including mirrors, lenses, polarizer, beam splitter, spatial light modulator, and the multimode fiber used for speckle acquisition.}
\label{setup}
\end{figure}

The experimental setup of this work is shown in Fig.~\ref{setup}. A 633~nm laser diode~(Eagleyard GC-02940), driven by a Thorlabs CLD1015 controller, serves as the light source. The beam is directed by mirrors and collimated using a telescope made from two lenses with focal lengths of 500~mm and 100~mm. A polarizer placed after the telescope ensures proper alignment with the HOLOEYE Pluto 2.0 spatial light modulator. The polarized beam then passes through a 50/50 beam splitter~(BS), where one part is transmitted toward the SLM and the other is blocked. The SLM reflects the phase-modulated beam, which re-enters the BS and is focused by another lens onto a fiber collimator that couples the light into a multimode fiber. The MMF used has a 400~$\mu$m core diameter, a numerical aperture of 0.22, and a length of 1~m.

For the data augmentation experiments, we use the OrganAMNIST dataset, consisting of 58,830 grayscale medical images. This dataset is chosen for its complexity and scale, allowing us to examine whether synthetic data generation can effectively reduce experimental acquisition time. To simulate data scarcity, smaller subsets of the dataset are used and augmented accordingly. Each image is encoded onto the laser beam via the SLM, which modulates the phase of the incident light. As the beam propagates through the MMF, modal interference transforms the encoded image into a complex speckle pattern. The resulting speckles emerging from the fiber's distal end are imaged onto a Thorlabs DCC1545M CMOS camera via an additional lens, and the captured intensity patterns are stored on a computer to form the output dataset.
\section{Standard data augmentation}  \label{sec:aug}
\begin{figure*}[htbp]
\centering\includegraphics[width=4.8in]{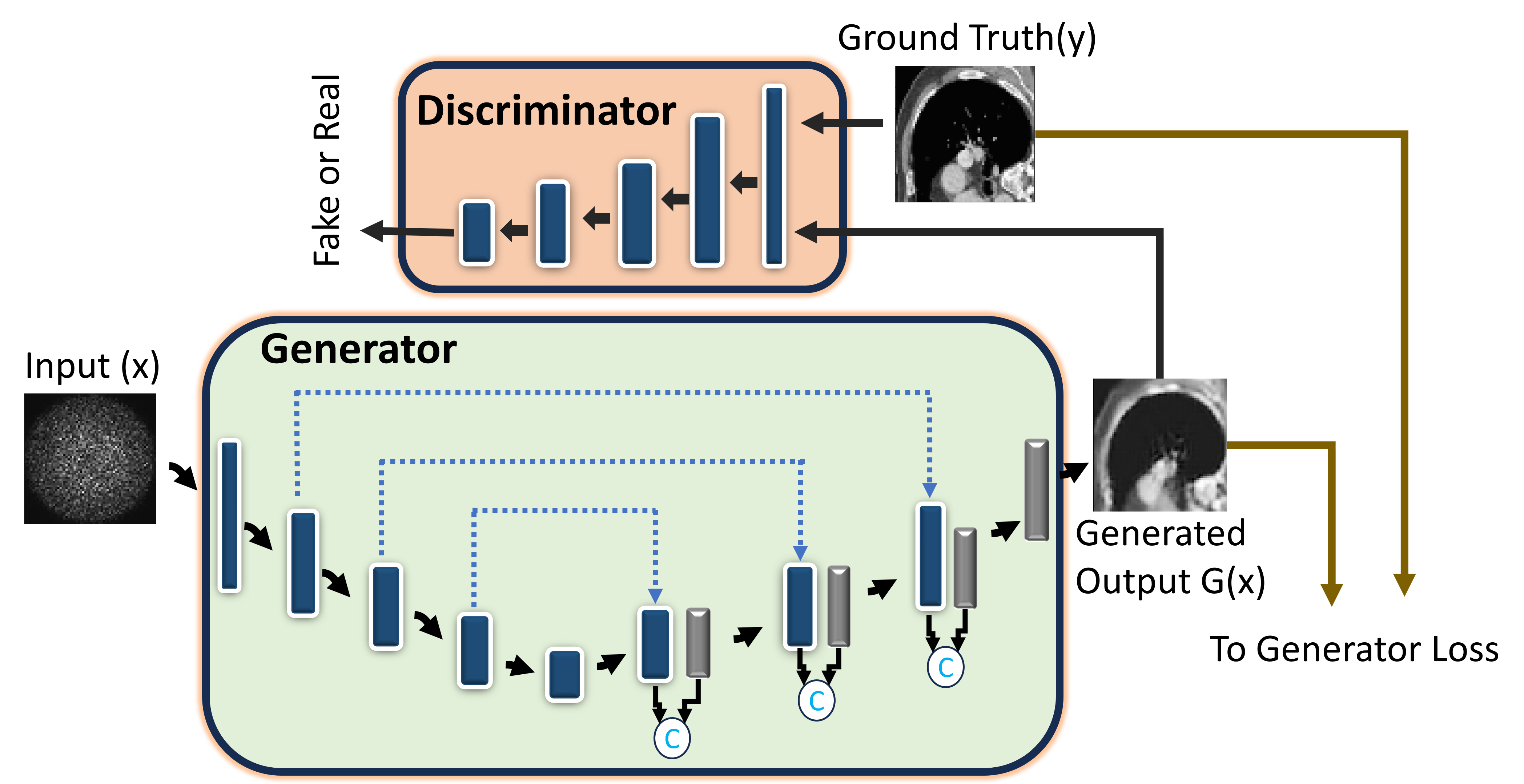}
\caption{The architecture of conditional generative adversarial network}
\label{cgan}
\end{figure*}

After recording 58,830 organ–speckle pairs, 5\% randomly selected samples are reserved for validation, and 10\% unseen samples are held out as a test set that remains unused throughout all augmentation trainings. To examine the effect of standard data augmentation techniques in multimode fiber imaging under limited-data conditions, only 5,000 samples from the remaining data are used for training.  We first train a conditional generative adversarial network, shown in Fig.~\ref{cgan}, where the generator learns to reconstruct the target image~(organ) from its corresponding speckle pattern. At the same time, the discriminator distinguishes between real and reconstructed images. In this adversarial setup, the generator progressively learns to produce outputs that the discriminator cannot differentiate from the true labels, thereby improving reconstruction quality over time.

The generator loss function combines adversarial, perceptual, and structural similarity losses. The adversarial component enforces realism in the generated outputs, while the perceptual loss captures high-level semantic information by comparing deep features extracted from a pretrained VGG network~\cite{johnson2016perceptual}. The SSIM term preserves essential textures and spatial relationships by comparing luminance, contrast, and structure between the reconstructed and reference images~\cite{wang2004image}. The overall generator loss is expressed as:
\begin{equation}
\mathcal{L}_{G} = \lambda_{adv}\mathcal{L}_{adv} + \lambda_{perc}\mathcal{L}_{perc} + \lambda_{ssim}(1 - SSIM(G(x), y)),
\end{equation}
where $\lambda_{adv}$, $\lambda_{perc}$, and $\lambda_{ssim}$ are weighting factors for the corresponding loss terms.
The discriminator loss, which trains the model to distinguish between real and generated samples and utilizes binary cross-entropy, is defined as:
\begin{equation}
\mathcal{L}_{\text{D}} =   \frac{1}{2} \, \text{bce}(D(G(x), x), 0)+\frac{1}{2} \, \text{bce}(D(y, x), 1)
\end{equation}

\begin{figure*}[htbp]
\centering\includegraphics[width=4.8in]{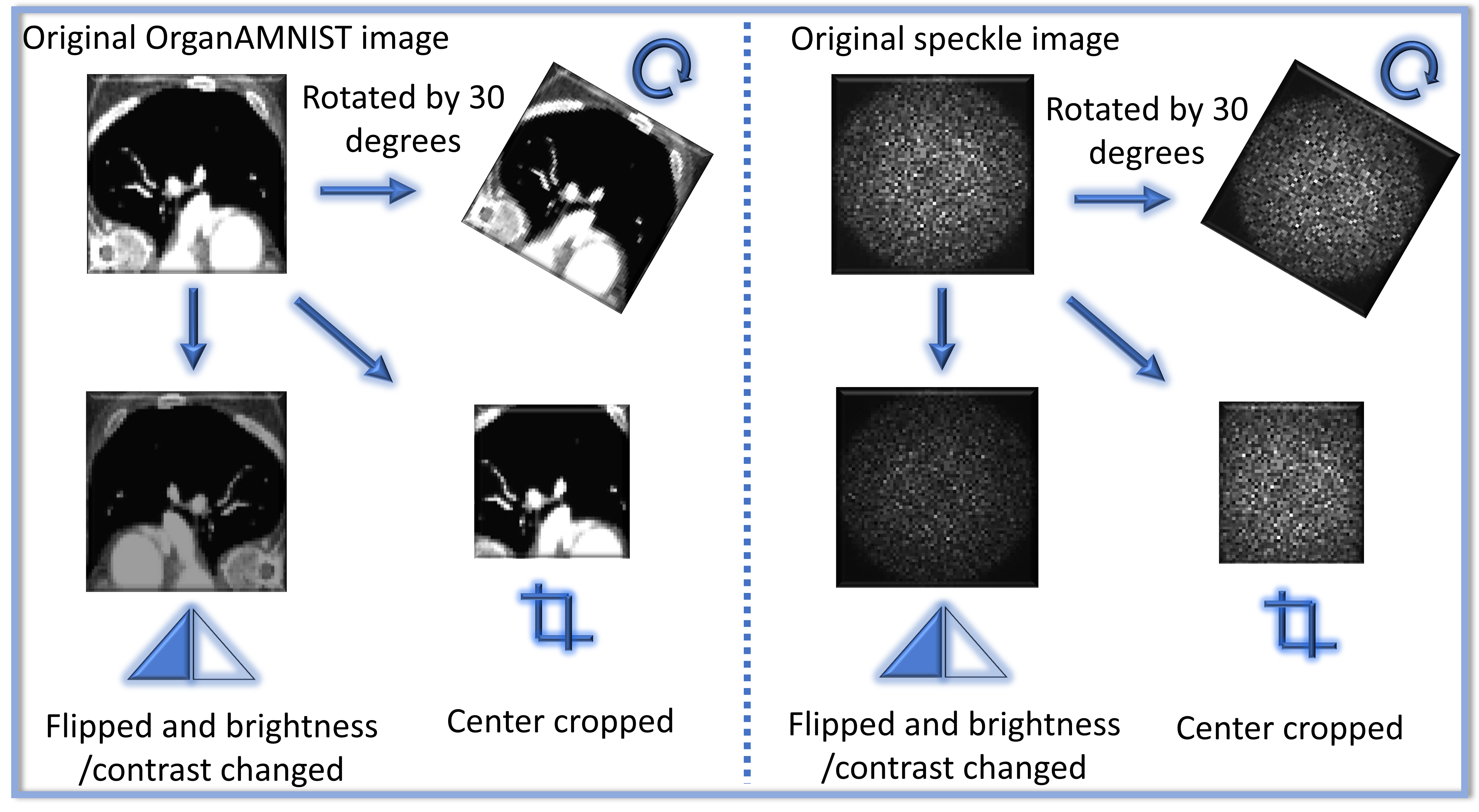}
\caption{ Illustration of standard augmentation techniques applied to multimode fiber imaging. Simple transformations such as rotation, flipping, cropping, and brightness changes alter the input organ images and their corresponding speckle patterns.}
\label{aug}
\end{figure*}

\begin{figure*}[htbp]
\centering\includegraphics[width=4.8in]{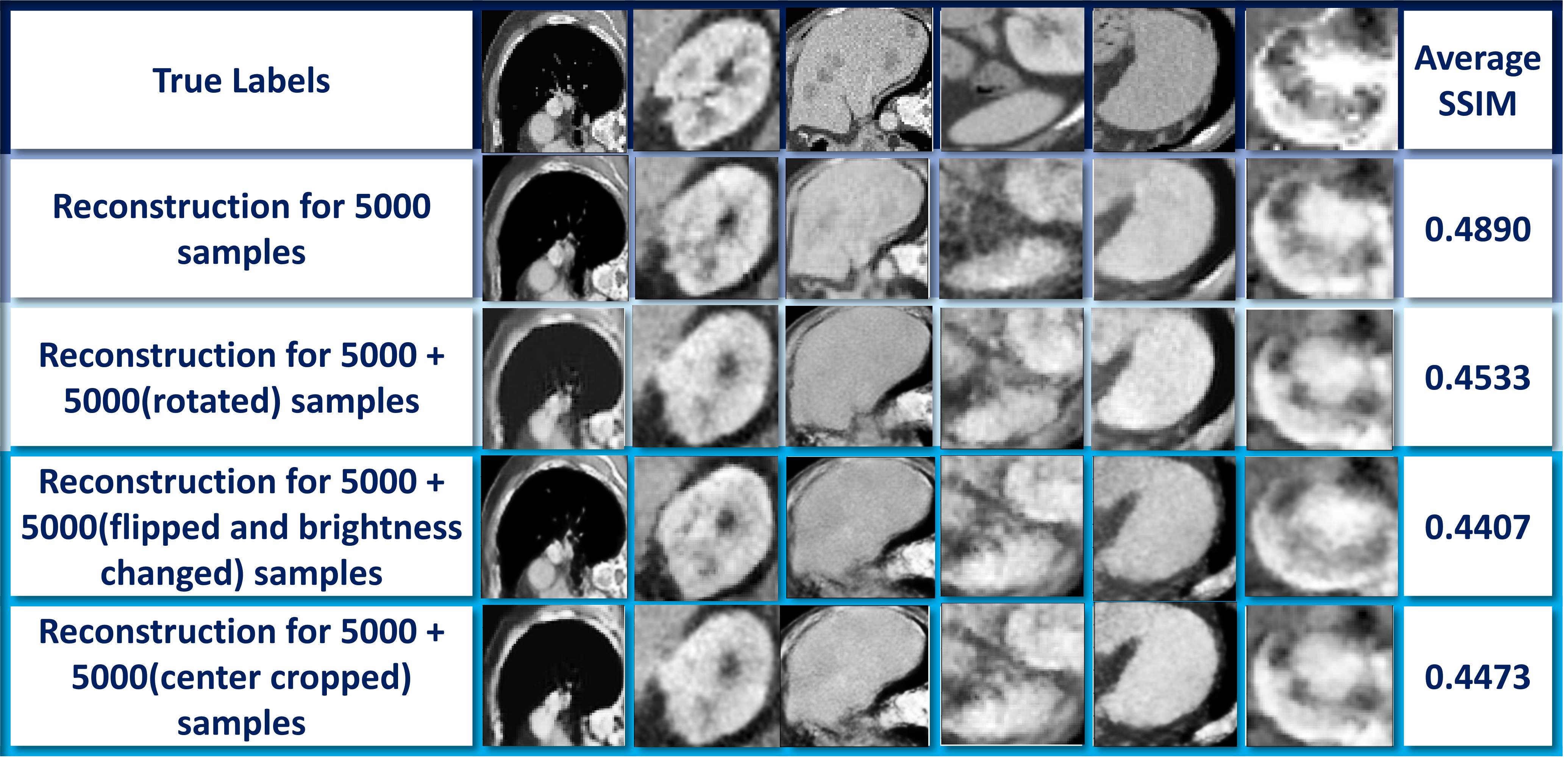}
\caption{Reconstruction results for standard augmentation experiments. The model is trained on 5,000 original and augmented datasets. The reconstructed images and their average SSIM values indicate that conventional digital augmentations degrade image reconstruction quality in MMF imaging.}
\label{res}
\end{figure*}

After training on 5,000 samples, the model reconstructs the 5,883 unseen test images, achieving an average SSIM of 0.4890. To assess the influence of traditional augmentation, we apply simple image transformations using PyTorch, including rotation by 30~degrees, horizontal flipping with brightness adjustment, and center cropping, to both the labels and their corresponding speckles. These transformations are depicted in Fig.~\ref{aug}.In this way, we form three augmented datasets of 10,000 samples each, which are used individually to train the network. Interestingly, the results show that the average SSIM on the test data decreases when trained with these augmented datasets, as shown in Fig.~\ref{res}.
This reduction occurs because simple digital transformations cannot be consistently applied to both the input (label) and output (speckle) images in MMF imaging. The physics of light propagation governs the relationship between an input image and its corresponding speckle pattern through the multimode fiber. The output speckle intensity $I_{out}(r)$ at a position $r$ is the squared magnitude of the total electric field, which is a coherent superposition of all $m$ guided modes:
\begin{equation}
I_{out}(r) = \left| E_{out}(r) \right|^2 = \left| \sum_{m=1}^{M} c_m \psi_m(r) e^{i\beta_m L} \right|^2,
\label{eq:speckle_formation}
\end{equation}
where $\psi_m(r)$ is the spatial profile of the $m$-th mode, $\beta_m$ is its propagation constant, $L$ is the fiber length, and $c_m$ is cross-correlation between the input field $E_{in}$ (at the $r'$ plane) and the modal field $\psi_m(r')$:
\begin{equation}
c_m = \int E_{in}(r') \psi_m^*(r') \, dr'.
\label{eq:overlap_integral}
\end{equation}
When a digital transformation such as flipping or rotation is applied to $E_{in}$, it modifies the correlation in Eq.~\ref{eq:overlap_integral}, resulting in a completely different set of $c_m$ coefficients. Consequently, the new speckle pattern that physically emerges from the fiber will differ fundamentally from a simple flipped or rotated version of the original speckle. This mismatch disrupts the underlying physical mapping between the input image and its corresponding speckle pattern, causing the network to learn incorrect relationships and reducing reconstruction accuracy.
\section{Data augmentation using CGAN}
\begin{figure*}[htpb]
\centering\includegraphics[width=5.2in]{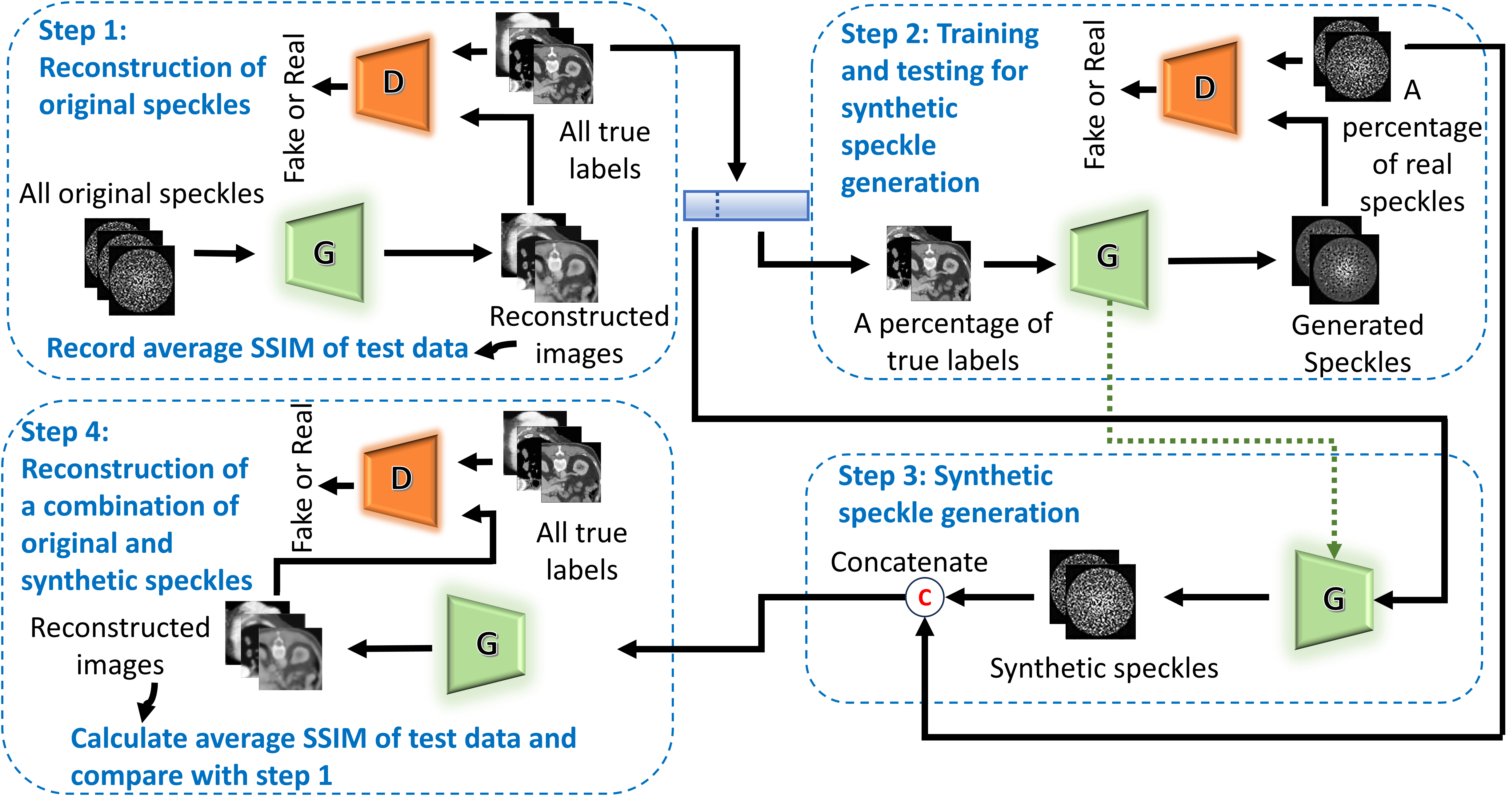}
\caption{Stepwise workflow of the CGAN-based data augmentation process for multimode fiber imaging. The figure outlines how the model is first trained on real speckles, then used to generate synthetic speckles from organ images, and finally retrained on a mixed dataset of real and synthetic samples to evaluate the impact on reconstruction quality.}
\label{steps}
\end{figure*}

\begin{figure*}[htbp]
\centering\includegraphics[width=5.1in]{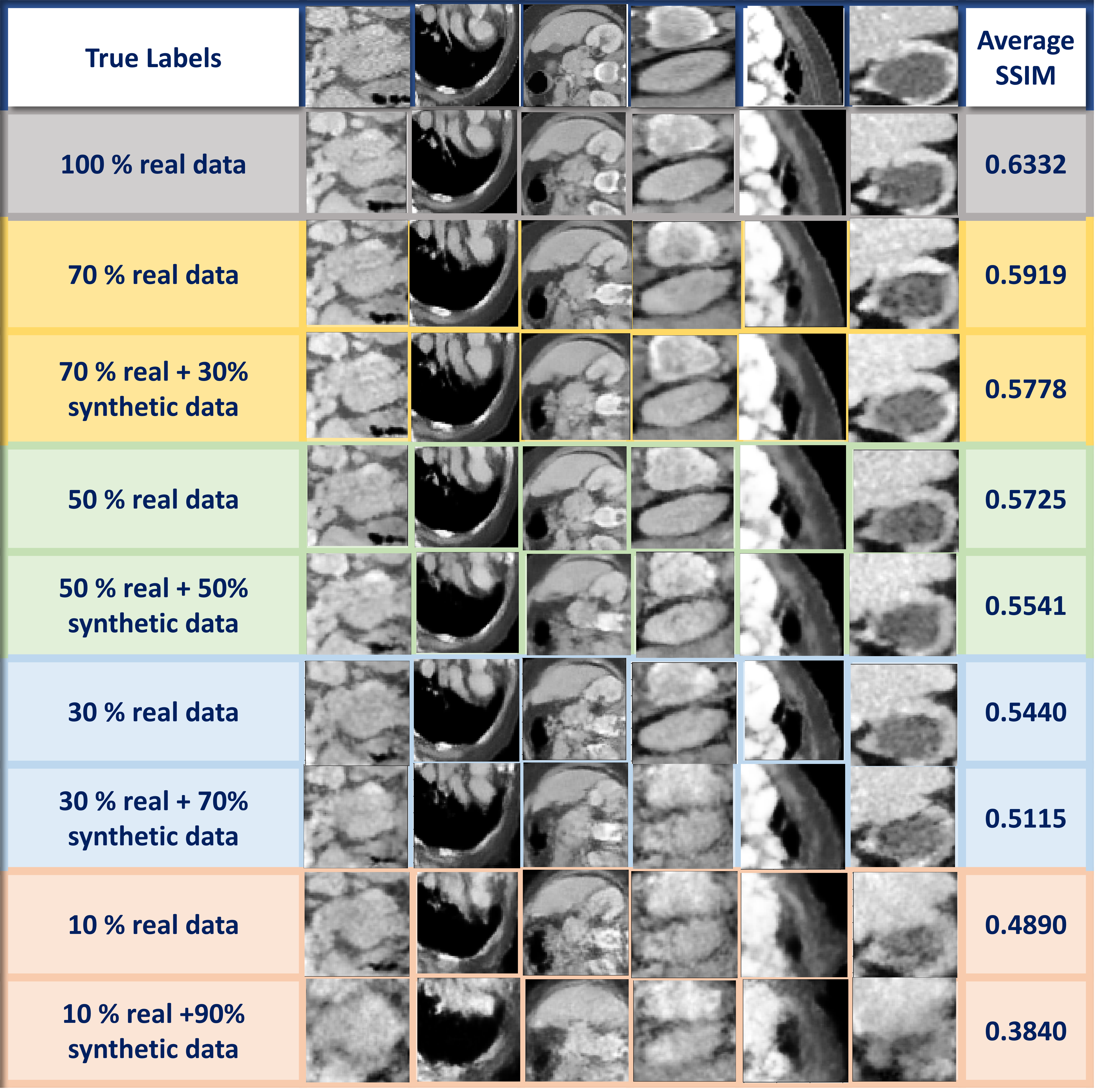}
\caption{Reconstruction performance of the deep learning model when trained on various combinations of real and CGAN-generated synthetic data.}
\label{results}
\end{figure*}

Acquiring 58,830 organ–speckle pairs requires nearly 25 hours of continuous operation. When datasets are recorded for multiple fiber configurations to achieve robustness in multimode fiber imaging, the total acquisition time increases proportionally. Maintaining constant experimental conditions, such as temperature, vibration, and alignment, over such extended durations is highly challenging. While collecting a smaller dataset can reduce time, it will also compromise reconstruction accuracy and model generalization. To address this, we investigate the use of a conditional Generative Adversarial Network to synthesize realistic speckle patterns to reduce physical acquisition time. Our experimental workflow is illustrated in Fig.~\ref{steps}.
We first create a baseline by training our reconstruction network on the whole 50,000 real-sample training set, achieving an average test SSIM of 0.6332, as shown in the top row of Fig.~\ref{results}.
We then implement our hypothesis using partial data. The experiment proceeds in four steps, using a 70\% real-data split as an example: \textbf{(Step 1):} The reconstruction network is trained on only 70\% of the real data (35,000 samples). This gives a reduced average SSIM of 0.5919. This is the new baseline using fewer samples.\textbf{(Step 2):} A CGAN is trained on this same 70\% dataset, but the process is now opposite. It now learns to generate synthetic speckles from their corresponding organ labels.\textbf{(Step 3):} The trained CGAN generator is used to generate speckles for the remaining 30\% of organ images. These synthetic speckles are concatenated with the original 70\% real speckles to create a full-size (50,000-sample) hybrid training set. \textbf{(Step 4):} The reconstruction network is trained from scratch on this new hybrid (70\% real + 30\% synthetic) dataset.

\begin{figure*}[htbp]
\centering\includegraphics[width=5.1in]{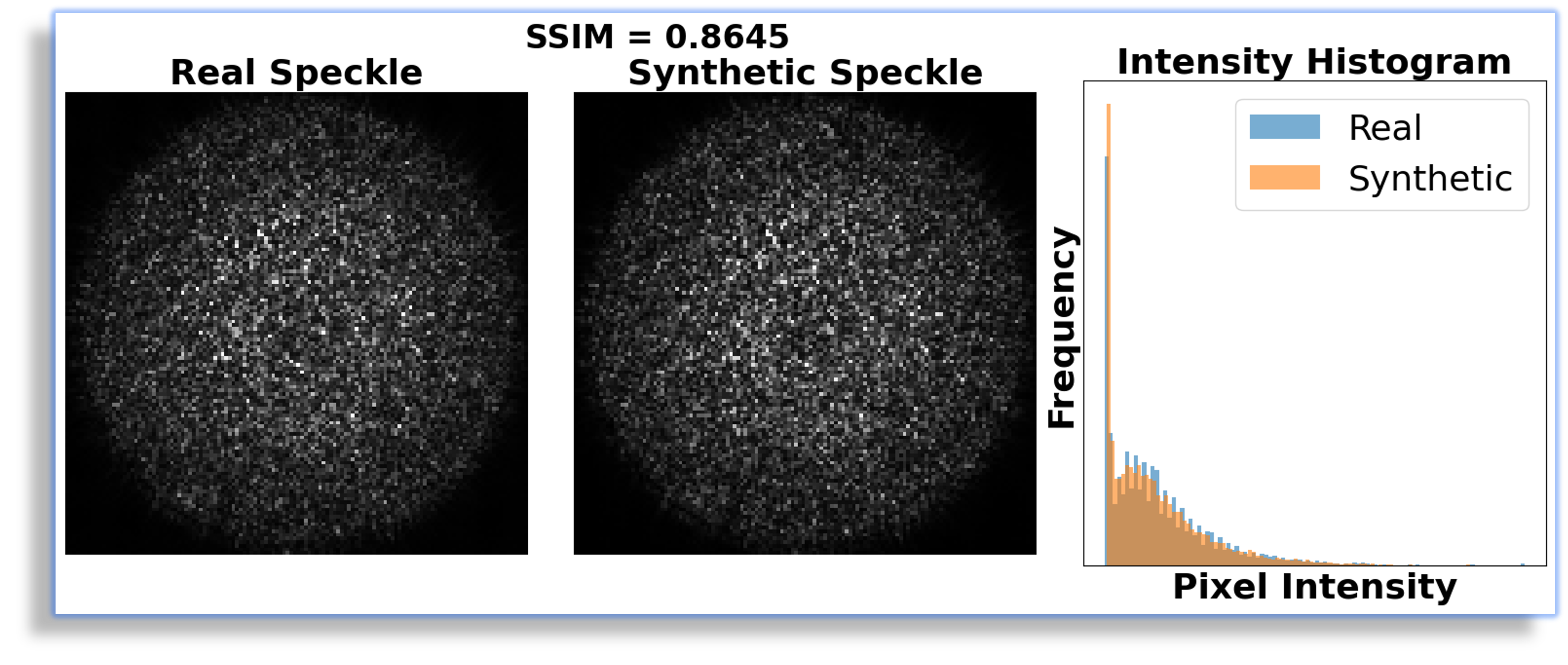}
\caption{Real and generated speckles for test data. It shows a synthetic speckle with a relatively high SSIM (0.8645); however, its intensity histogram still reveals a noticeable difference from the actual distribution.}
\label{speckle}
\end{figure*}

We hypothesize that if the CGAN produces realistic speckles, the SSIM from Step 4 should recover towards the 100\% real-data baseline (0.6332), or at least surpass the 70\%-only baseline (0.5919). However, after training on 70\% real and 30\% synthetic data, we achieve an average SSIM of 0.5778, which is lower than the 70\%-only baseline. As summarized in Fig.~\ref{results}, we repeat this entire process for 50\%, 30\%, and 10\% real-data splits. In every case, adding CGAN-generated synthetic data degrades reconstruction performance compared to using the partial real dataset alone. This degradation becomes more severe as the proportion of synthetic data increases, indicating that, rather than being helpful, the CGAN-generated synthetic speckles further degrade reconstruction performance. This deterioration occurs because the CGAN fails to accurately capture the high-dimensional statistical distribution of MMF speckles. For the 70\% training data scenario, the average SSIM of the generated speckles for unseen samples is 0.8821. For different percentages of real training data (50\%, 30\%, and 10\%), only a minor drop in SSIM is observed (from 0.8821 to 0.8614). This relatively consistent, high SSIM initially suggests the CGAN is performing well. However, this is misleading, as the CGAN tends to produce speckles that are visually similar in overall structure but fail to accurately reproduce the fine pixel-level intensity distribution.  This is evident in Fig.~\ref{speckle}, which shows that while a synthetic speckle may have a high SSIM, its pixel intensity histogram is visibly different. The CGAN fails to reproduce the precise probability distribution of high-intensity areas and dark, zero-intensity, regions. As the percentage of real data used to train the CGAN decreases, its ability to learn the actual data distribution worsens, leading to even lower-quality synthetic samples. This can be seen in the case of only 10\% real and 90\% synthetic speckles, where the average SSIM is too low. This limitation has also been observed in other deep learning tasks, where GAN-based data augmentation can introduce statistical bias and reduce performance~\cite{bissoto2018skin,jindal2024bias,ali2023leveraging}. In MMF imaging, this failure is further exacerbated by the nonlinear nature of light propagation~\cite{caramazza2019transmission}, highlighting the importance of physically informed generative models for producing realistic synthetic speckle data.

\section{Physical Data Augmentation through Experimental Acquisition}
We have shown that standard digital augmentation fails in MMF imaging and that synthetic speckle generation via CGANs cannot replace experimental measurements. We now introduce a new physical data augmentation strategy explicitly designed for multimode fiber systems. In our method, only the label images (organ images) are digitally transformed, and their corresponding speckle patterns are experimentally captured by displaying these transformed images on the spatial light modulator and transmitting them through the multimode fiber. It preserves the accurate optical correspondence between the input field and its recorded speckle pattern.

\begin{figure}[h]
\centering
\includegraphics[width=\textwidth]{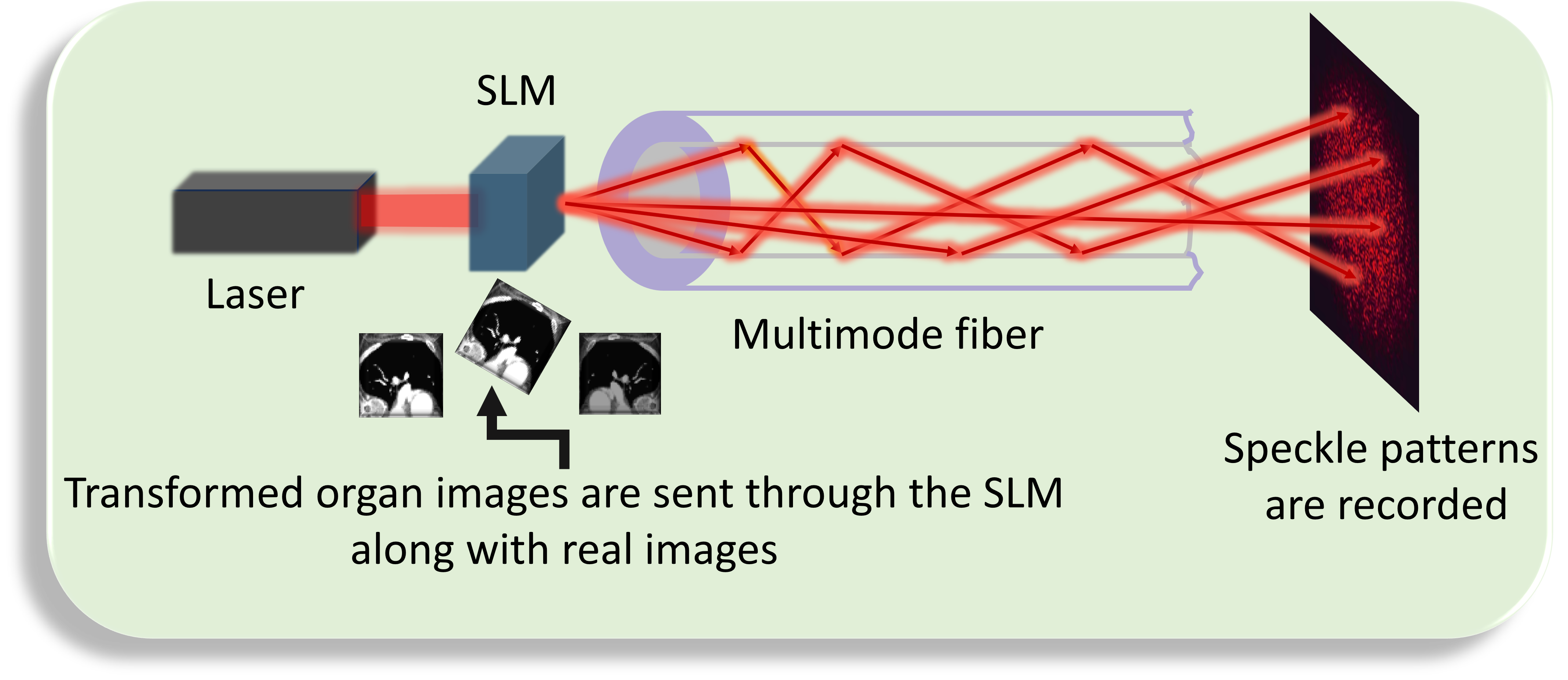}
\caption{Experimental schematics for the physical data augmentation. Transformed organ images are displayed on an SLM, coupled into the multimode fiber, and the resulting output speckle patterns are recorded.}
\label{setup_aug}
\end{figure}

\begin{figure}[h]
\centering
\includegraphics[width=\textwidth]{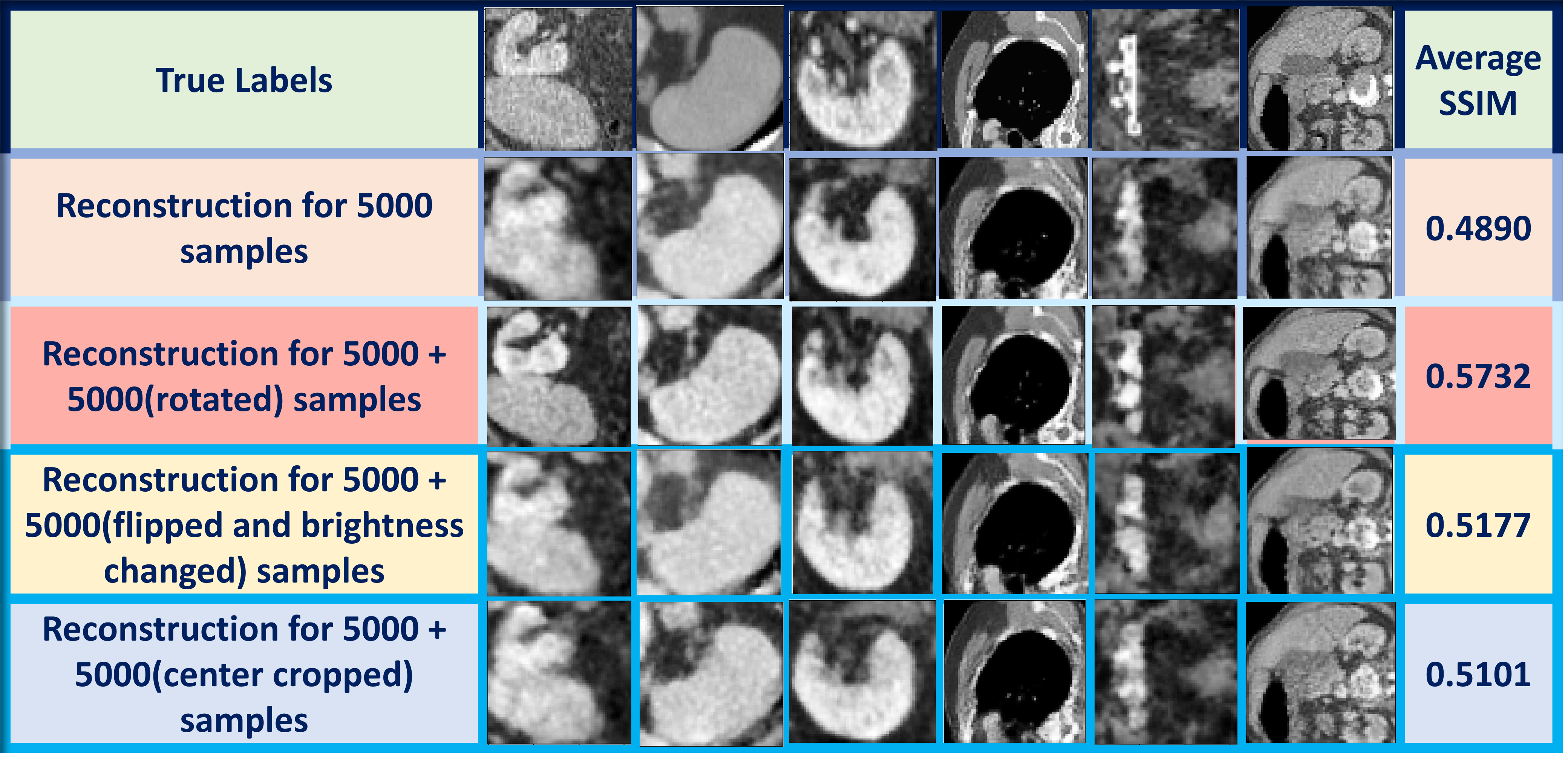}
\caption{Comparison of reconstruction quality for different physical data augmentation strategies.}
\label{results2}
\end{figure}

As shown in Section~\ref{sec:aug}, training the CGAN model on 5,000 real organ–speckle pairs gives an average SSIM of 0.4890 on 5,883 unseen test samples. In our proposed augmentation scheme, we first rotate each of these 5000 organ images by 30~degrees and add the transformed samples to the original 5,000, creating a dataset of 10,000 organ images. Each of these images, along with the test set, is projected onto the SLM, transmitted through the MMF, and the corresponding output speckles are recorded, as illustrated in Fig.~\ref{setup_aug}. When the CGAN model is trained on this augmented dataset, the average test SSIM increases by 17\% to 0.5732. To further evaluate the impact of this approach, we perform additional transformations, including center cropping and horizontal flipping combined with brightness modification. The test SSIM values improve to 0.5101(4.3\%) and 0.5177(5.8\%), respectively, as shown in Fig.~\ref{results2}. These results demonstrate that the proposed physical augmentation method enhances generalization and reconstruction accuracy under limited data conditions, confirming its effectiveness for different transformation types.

The fact that SSIM is more improved in the rotation-based experiment than in other transformations is attributable to the physics of light propagation in multimode fibers. When an image is displayed on the SLM, it modulates the phase values of the incident field. This modulated field couples into the fiber modes according to Eq.~\ref{eq:overlap_integral}. A simple brightness change, for instance, scales all modal excitation coefficients uniformly and changes the overall speckle intensity with little alteration in mode coupling. Similarly, a horizontal flip is a relatively simple change in spatial information. Center cropping mainly reduces the spatial dimension of the input with a slight alteration in its angular structure. The coupling to higher-order modes is reduced, but the relative distribution among excited modes remains similar.

In contrast, a $30^{\circ}$ rotation represents a pronounced spatial transformation. By reorienting the input field, $E_{in}(r') \rightarrow E_{in}(R_{30}r')$, it results in a complex linear combination of the guided modes, yielding a substantially different set of excitation coefficients $c_m$. Consequently, a diverse speckle data set is produced, which compels the reconstruction model to learn a more generalized and robust mapping. These results demonstrate that physical augmentation can substantially improve deep learning performance in MMF imaging when training data is limited. The key distinction from standard augmentation is that we preserve the true physical mapping between input transformations and output speckle patterns by collecting experimental data. While this approach requires additional measurement time, it enables effective data augmentation in scenarios where the underlying physics cannot be accurately simulated and where some extra time can be tolerated to achieve better model performance.
\section{Conclusion}
We have, for the first time, systematically demonstrated that traditional augmentation and generative methods fail to improve image reconstruction through multimode fibers. Unlike conventional imaging tasks, where simple digital transformations augment the dataset, multimode fiber imaging is governed by complex optical interference that cannot be reproduced by synthetic or digitally altered data alone. This finding emphasizes the need to integrate the physical properties of light propagation into the data preparation process. We have proposed and implemented a physical augmentation strategy, in which only organ images are digitally transformed, while their corresponding speckles are experimentally recorded. This method yields a significant improvement, increasing reconstruction SSIM by 17\%. These findings are important because they redefine how data augmentation should be approached for optically complex imaging systems. Rather than viewing augmentation purely as a digital operation, our results suggest that physically consistent data generation can substantially enhance model robustness and accuracy under limited-data conditions. Although this approach increases acquisition time, it provides a viable pathway for creating diverse and realistic datasets without compromising optical integrity. It can be employed in cases where high accuracy is required, even with limited data, and additional data acquisition can be tolerated.  Future work will focus on developing generative models that accurately simulate the full statistical distribution of speckle patterns using physics-informed learning to reduce the need for extensive physical data collection. The principles of our work not only apply to MMF-based medical endoscopy but can be extended to other complex optical systems where physics governs light propagation. These include computational microscopy and imaging through other scattering media, such as biological tissue, fog, and haze, hence offering a new direction for effective data augmentation in domains that remain unaddressed.

\section*{Funding}
This work is funded by the Syed Babar Ali Research Award (SBARA) (GRA-0085) for the year 2025–2026.

\section*{Disclosures}
The authors declare no conflicts of interest.

\section*{Data Availability}
Codes and data underlying the results presented in this paper are not publicly available at this time, but may be obtained from the authors upon reasonable request.

%%%%%%%%%% If using BibTeX:
\bibliographystyle{unsrt}
\bibliography{synthetic}

@article{papadopoulos2012focusing,
  title={Focusing and scanning light through a multimode optical fiber using digital phase conjugation},
  author={Papadopoulos, Ioannis N and Farahi, Salma and Moser, Christophe and Psaltis, Demetri},
  journal={Optics express},
  volume={20},
  number={10},
  pages={10583--10590},
  year={2012},
  publisher={Optica Publishing Group}
}

@article{papadopoulos2013high,
  title={High-resolution, lensless endoscope based on digital scanning through a multimode optical fiber},
  author={Papadopoulos, Ioannis N and Farahi, Salma and Moser, Christophe and Psaltis, Demetri},
  journal={Biomedical optics express},
  volume={4},
  number={2},
  pages={260--270},
  year={2013},
  publisher={Optica Publishing Group}
}

@article{popoff2010measuring,
  title={Measuring the transmission matrix in optics: an approach to the study and control of light propagation in disordered media},
  author={Popoff, S{\'e}bastien M and Lerosey, Geoffroy and Carminati, R{\'e}mi and Fink, Mathias and Boccara, Albert Claude and Gigan, Sylvain},
  journal={Physical review letters},
  volume={104},
  number={10},
  pages={100601},
  year={2010},
  publisher={APS}
}

@inproceedings{akbulut2013measurements,
  title={Measurements on the optical transmission matrices of strongly scattering nanowire layers},
  author={Akbulut, Duygu and Strudley, Tom and Bertolotti, Jacopo and Zehender, Tilman and Bakkers, Erik PAM and Lagendijk, Ad and Vos, Willem L and Muskens, Otto L and Mosk, Allard P},
  booktitle={International Quantum Electronics Conference},
  pages={IH\_P\_19},
  year={2013},
  organization={Optica Publishing Group}
}

@article{borhani2018learning,
  title={Learning to see through multimode fibers},
  author={Borhani, Navid and Kakkava, Eirini and Moser, Christophe and Psaltis, Demetri},
  journal={Optica},
  volume={5},
  number={8},
  pages={960--966},
  year={2018},
  publisher={Optical Society of America}
}

@article{rahmani2018multimode,
  title={Multimode optical fiber transmission with a deep learning network},
  author={Rahmani, Babak and Loterie, Damien and Konstantinou, Georgia and Psaltis, Demetri and Moser, Christophe},
  journal={Light: science \& applications},
  volume={7},
  number={1},
  pages={69},
  year={2018},
  publisher={Nature Publishing Group UK London}
}

@article{zhu2021image,
  title={Image reconstruction through a multimode fiber with a simple neural network architecture},
  author={Zhu, Changyan and Chan, Eng Aik and Wang, You and Peng, Weina and Guo, Ruixiang and Zhang, Baile and Soci, Cesare and Chong, Yidong},
  journal={Scientific reports},
  volume={11},
  number={1},
  pages={896},
  year={2021},
  publisher={Nature Publishing Group UK London}
}

@article{yu2021high,
  title={High-speed multimode fiber imaging system based on conditional generative adversarial network},
  author={Yu, Zhenming and Ju, Zhenyu and Zhang, Xinlei and Meng, Ziyi and Yin, Feifei and Xu, Kun},
  journal={Chinese Optics Letters},
  volume={19},
  number={8},
  pages={081101},
  year={2021},
  publisher={Chinese Optical Society}
}

@article{wang2022upconversion,
  title={Upconversion imaging through multimode fibers based on deep learning},
  author={Wang, Xiaoyan and Wang, Zhiyuan and Luo, Songjie and Chen, Ziyang and Pu, Jixiong},
  journal={Optik},
  volume={264},
  pages={169444},
  year={2022},
  publisher={Elsevier}
}

@inproceedings{isola2017image,
  title={Image-to-image translation with conditional adversarial networks},
  author={Isola, Phillip and Zhu, Jun-Yan and Zhou, Tinghui and Efros, Alexei A},
  booktitle={Proceedings of the IEEE conference on computer vision and pattern recognition},
  pages={1125--1134},
  year={2017}
}

@inproceedings{kremp2023neural,
  title={Neural-network-based multimode fiber imaging and position sensing under thermal perturbations},
  author={Kremp, Tristan and Bagley, Nicholas and Lamb, Erin S and Westbrook, Paul S and DiGiovanni, David J},
  booktitle={Adaptive Optics and Wavefront Control for Biological Systems IX},
  volume={12388},
  pages={35--48},
  year={2023},
  organization={SPIE}
}

@article{abdulaziz2023robust,
  title={Robust real-time imaging through flexible multimode fibers},
  author={Abdulaziz, Abdullah and Mekhail, Simon Peter and Altmann, Yoann and Padgett, Miles J and McLaughlin, Stephen},
  journal={Scientific Reports},
  volume={13},
  number={1},
  pages={11371},
  year={2023},
  publisher={Nature Publishing Group UK London}
}

@inproceedings{maqbool2024application,
  title={Application of conditional generative adversarial networks toward time-efficient and high-fidelity imaging via multimode fibers},
  author={Maqbool, Jawaria and Hassan, Syed Talal and Cheema, M Imran},
  booktitle={AI and Optical Data Sciences V},
  volume={12903},
  pages={69--73},
  year={2024},
  organization={SPIE}
}

@article{feng2025high,
  title={High-fidelity image reconstruction in multimode fiber imaging through the MITM-Unet framework},
  author={Feng, Zefeng and Yue, Zengqi and Zhou, Wei and Xu, Baoteng and Liu, Jialin and Sun, Jiawei and Xiong, Daxi and Yang, Xibin},
  journal={Optics Express},
  volume={33},
  number={3},
  pages={5866--5876},
  year={2025},
  publisher={Optica Publishing Group}
}

@article{maqbool2024towards,
  title={Towards optimal multimode fiber imaging by leveraging input polarization and deep learning},
  author={Maqbool, Jawaria and Hasan, Syed Talal and Cheema, M Imran},
  journal={Optical Fiber Technology},
  volume={87},
  pages={103896},
  year={2024},
  publisher={Elsevier}
}

@article{caramazza2019transmission,
  title={Transmission of natural scene images through a multimode fibre},
  author={Caramazza, Piergiorgio and Moran, Ois{\'\i}n and Murray-Smith, Roderick and Faccio, Daniele},
  journal={Nature communications},
  volume={10},
  number={1},
  pages={2029},
  year={2019},
  publisher={Nature Publishing Group UK London}
}

@inproceedings{bissoto2018skin,
  title={Skin lesion synthesis with generative adversarial networks},
  author={Bissoto, Alceu and Perez, F{\'a}bio and Valle, Eduardo and Avila, Sandra},
  booktitle={OR 2.0 Context-Aware Operating Theaters, Computer Assisted Robotic Endoscopy, Clinical Image-Based Procedures, and Skin Image Analysis: First International Workshop, OR 2.0 2018, 5th International Workshop, CARE 2018, 7th International Workshop, CLIP 2018, Third International Workshop, ISIC 2018, Held in Conjunction with MICCAI 2018, Granada, Spain, September 16 and 20, 2018, Proceedings 5},
  pages={294--302},
  year={2018},
  organization={Springer}
}

@article{krizhevsky2012imagenet,
  title={Imagenet classification with deep convolutional neural networks},
  author={Krizhevsky, Alex and Sutskever, Ilya and Hinton, Geoffrey E},
  journal={Advances in neural information processing systems},
  volume={25},
  year={2012}
}

@article{kumar2024deep,
  title={Deep learning for hyperspectral image classification: A survey},
  author={Kumar, Vinod and Singh, Ravi Shankar and Rambabu, Medara and Dua, Yaman},
  journal={Computer Science Review},
  volume={53},
  pages={100658},
  year={2024},
  publisher={Elsevier}
}

@article{brar2025image,
  title={Image segmentation review: Theoretical background and recent advances},
  author={Brar, Khushmeen Kaur and Goyal, Bhawna and Dogra, Ayush and Mustafa, Mohammed Ahmed and Majumdar, Rana and Alkhayyat, Ahmed and Kukreja, Vinay},
  journal={Information Fusion},
  volume={114},
  pages={102608},
  year={2025},
  publisher={Elsevier}
}

@inproceedings{liew2017regional,
  title={Regional interactive image segmentation networks},
  author={Liew, JunHao and Wei, Yunchao and Xiong, Wei and Ong, Sim-Heng and Feng, Jiashi},
  booktitle={2017 IEEE international conference on computer vision (ICCV)},
  pages={2746--2754},
  year={2017},
  organization={IEEE}
}

@article{aung2024review,
  title={A review of lidar-based 3d object detection via deep learning approaches towards robust connected and autonomous vehicles},
  author={Aung, Nang Htet Htet and Sangwongngam, Paramin and Jintamethasawat, Rungroj and Shah, Shashi and Wuttisittikulkij, Lunchakorn},
  journal={IEEE Transactions on Intelligent Vehicles},
  year={2024},
  publisher={IEEE}
}

@article{lee2024holosr,
  title={HoloSR: deep learning-based super-resolution for real-time high-resolution computer-generated holograms},
  author={Lee, Siwoo and Nam, Seung-Woo and Lee, Juhyun and Jeong, Yoonchan and Lee, Byoungho},
  journal={Optics Express},
  volume={32},
  number={7},
  pages={11107--11122},
  year={2024},
  publisher={Optica Publishing Group}
}

@article{su2025review,
  title={A review of deep-learning-based super-resolution: From methods to applications},
  author={Su, Hu and Li, Ying and Xu, Yifan and Fu, Xiang and Liu, Song},
  journal={Pattern Recognition},
  volume={157},
  pages={110935},
  year={2025},
  publisher={Elsevier}
}

@inproceedings{sun2017revisiting,
  title={Revisiting unreasonable effectiveness of data in deep learning era},
  author={Sun, Chen and Shrivastava, Abhinav and Singh, Saurabh and Gupta, Abhinav},
  booktitle={Proceedings of the IEEE international conference on computer vision},
  pages={843--852},
  year={2017}
}

@article{russakovsky2015imagenet,
  title={Imagenet large scale visual recognition challenge},
  author={Russakovsky, Olga and Deng, Jia and Su, Hao and Krause, Jonathan and Satheesh, Sanjeev and Ma, Sean and Huang, Zhiheng and Karpathy, Andrej and Khosla, Aditya and Bernstein, Michael and others},
  journal={International journal of computer vision},
  volume={115},
  number={3},
  pages={211--252},
  year={2015},
  publisher={Springer}
}

@article{sharma2025addressing,
  title={Addressing class imbalance in remote sensing using deep learning approaches: a systematic literature review},
  author={Sharma, Shweta and Gosain, Anjana},
  journal={Evolutionary Intelligence},
  volume={18},
  number={1},
  pages={23},
  year={2025},
  publisher={Springer}
}

@inproceedings{cieslak2024generating,
  title={Generating diverse agricultural data for vision-based farming applications},
  author={Cieslak, Mikolaj and Govindarajan, Umabharathi and Garcia, Alejandro and Chandrashekar, Anuradha and Hadrich, Torsten and Mendoza-Drosik, Aleksander and Michels, Dominik L and Pirk, Soren and Fu, Chia-Chun and Palubicki, Wojciech},
  booktitle={Proceedings of the IEEE/CVF Conference on Computer Vision and Pattern Recognition},
  pages={5422--5431},
  year={2024}
}

@article{meng2025tlstmf,
  title={TLSTMF-YOLO: Transfer learning and feature fusion network for earthquake-induced landslide detection in remote sensing images},
  author={Meng, Shaoqiang and Shi, Zhenming and Pirasteh, Saied and Ullo, Silvia Liberata and Peng, Ming and Zhou, Changshi and Goncalves, Wesley Nunes and Zhang, Limin},
  journal={IEEE Transactions on Geoscience and Remote Sensing},
  year={2025},
  publisher={IEEE}
}

@article{goceri2023medical,
  title={Medical image data augmentation: techniques, comparisons and interpretations},
  author={Goceri, Evgin},
  journal={Artificial intelligence review},
  volume={56},
  number={11},
  pages={12561--12605},
  year={2023},
  publisher={Springer}
}

@article{khalifa2022comprehensive,
  title={A comprehensive survey of recent trends in deep learning for digital images augmentation},
  author={Khalifa, Nour Eldeen and Loey, Mohamed and Mirjalili, Seyedali},
  journal={Artificial Intelligence Review},
  volume={55},
  number={3},
  pages={2351--2377},
  year={2022},
  publisher={Springer}
}

@article{de2022geometric,
  title={Geometric transformation-based data augmentation on defect classification of segmented images of semiconductor materials using a ResNet50 convolutional neural network},
  author={de la Rosa, Francisco L{\'o}pez and G{\'o}mez-Sirvent, Jos{\'e} L and S{\'a}nchez-Reolid, Roberto and Morales, Rafael and Fern{\'a}ndez-Caballero, Antonio},
  journal={Expert Systems with Applications},
  volume={206},
  pages={117731},
  year={2022},
  publisher={Elsevier}
}

@article{mumuni2024survey,
  title={A survey of synthetic data augmentation methods in computer vision},
  author={Mumuni, Alhassan and Mumuni, Fuseini and Gerrar, Nana Kobina},
  journal={arXiv preprint arXiv:2403.10075},
  year={2024}
}

@article{kirichenko2023understanding,
  title={Understanding the detrimental class-level effects of data augmentation},
  author={Kirichenko, Polina and Ibrahim, Mark and Balestriero, Randall and Bouchacourt, Diane and Vedantam, Shanmukha Ramakrishna and Firooz, Hamed and Wilson, Andrew G},
  journal={Advances in Neural Information Processing Systems},
  volume={36},
  pages={17498--17526},
  year={2023}
}

@inproceedings{gong2021keepaugment,
  title={Keepaugment: A simple information-preserving data augmentation approach},
  author={Gong, Chengyue and Wang, Dilin and Li, Meng and Chandra, Vikas and Liu, Qiang},
  booktitle={Proceedings of the IEEE/CVF conference on computer vision and pattern recognition},
  pages={1055--1064},
  year={2021}
}

@inproceedings{yue2022survey,
  title={Survey of image augmentation based on generative adversarial network},
  author={Yue, Fei and Zhang, Chao and Yuan, MingYang and Xu, Chen and Song, YaLin},
  booktitle={Journal of Physics: Conference Series},
  volume={2203},
  pages={012052},
  year={2022},
  organization={IOP Publishing}
}

@inproceedings{thakur2025ai,
  title={AI-Driven Synthetic Data Generation Using Conditional GANs to Mitigate Class Imbalance in COVID-19 X-Ray Imaging},
  author={Thakur, Amit and Mishra, Nilamadhab},
  booktitle={2025 International Conference on Advancements in Smart, Secure and Intelligent Computing (ASSIC)},
  pages={1--6},
  year={2025},
  organization={IEEE}
}

@article{ding2025improving,
  title={Improving imbalanced medical image classification through GAN-based data augmentation methods},
  author={Ding, Hongwei and Huang, Nana and Wu, Yaoxin and Cui, Xiaohui},
  journal={Pattern Recognition},
  volume={166},
  pages={111680},
  year={2025},
  publisher={Elsevier}
}

@article{paproki2024synthetic,
  title={Synthetic data for deep learning in computer vision \& medical imaging: A means to reduce data bias},
  author={Paproki, Anthony and Salvado, Olivier and Fookes, Clinton},
  journal={ACM Computing Surveys},
  volume={56},
  number={11},
  pages={1--37},
  year={2024},
  publisher={ACM New York, NY}
}

@inproceedings{ali2023leveraging,
  title={Leveraging GANs for data scarcity of COVID-19: Beyond the hype},
  author={Ali, Hazrat and Gr{\"o}nlund, Christer and Shah, Zubair},
  booktitle={Proceedings of the IEEE/CVF Conference on Computer Vision and Pattern Recognition},
  pages={659--667},
  year={2023}
}

@inproceedings{jindal2024bias,
  title={Bias inheritance and its amplification in GAN-based synthetic data augmentation for skin lesion classification},
  author={Jindal, Marut and Singh, Birmohan},
  booktitle={2024 3rd International Conference on Artificial Intelligence For Internet of Things (AIIoT)},
  pages={1--6},
  year={2024},
  organization={IEEE}
}

@inproceedings{maqbool2025deep,
  title={Deep learning-enabled imaging of abdominal organs through multimode fibers},
  author={Maqbool, Jawaria and Cheema, M Imran},
  booktitle={AI and Optical Data Sciences VI},
  volume={13375},
  pages={24--28},
  year={2025},
  organization={SPIE}
}

@inproceedings{johnson2016perceptual,
  title={Perceptual losses for real-time style transfer and super-resolution},
  author={Johnson, Justin and Alahi, Alexandre and Fei-Fei, Li},
  booktitle={European conference on computer vision},
  pages={694--711},
  year={2016},
  organization={Springer}
}

@article{wang2004image,
  title={Image quality assessment: from error visibility to structural similarity},
  author={Wang, Zhou and Bovik, Alan C and Sheikh, Hamid R and Simoncelli, Eero P},
  journal={IEEE transactions on image processing},
  volume={13},
  number={4},
  pages={600--612},
  year={2004},
  publisher={IEEE}
}

\end{document}